\documentclass[10pt]{article}

\usepackage{etoolbox}
\newtoggle{medphysjournal}
\togglefalse{medphysjournal}

\usepackage[utf8]{inputenc}
\usepackage[a4paper,lmargin=1.5in, rmargin=1.5in, footnotesep=1cm plus 4pt minus 4pt]{geometry}
\usepackage{setspace}
\usepackage{amsmath,amsthm,amssymb,esint}
\usepackage{mathtools, nccmath}
\allowdisplaybreaks
\usepackage{multirow}
\usepackage{sansmath}
\usepackage{pgfplots}
\pgfplotsset{compat = newest}
\usepackage{listings}
\usepackage{color}
\usepackage{placeins}
\usepackage{multicol}
\usepackage{multirow}
\usepackage{enumitem}
\usepackage{lipsum}
\usepackage{cases}
\usepackage{epstopdf}
\usepackage{float}
\usepackage{booktabs}
\usepackage{dirtytalk}
\usepackage{subcaption}
\usepackage{graphicx}

\usepackage{comment}

\usepackage[T1]{fontenc}
\usepackage{lmodern}
\usepackage{times}

\usepackage{sectsty}
\allsectionsfont{\sffamily\bfseries}
\subsubsectionfont{\sffamily\bfseries}

\usepackage[font=sf,labelfont=bf]{caption}

\usepackage[font=sf]{caption}

\newtheoremstyle{mystyle1}
  {}
  {}
  {\itshape}
  {}
  {\sffamily\bfseries}
  {{\sffamily\mdseries .}}
  { }
  {}
\theoremstyle{mystyle1}

\newtheoremstyle{mystyle2}
  {}
  {}
  {}
  {}
  {\itshape\sffamily}
  {\textsf{.}}
  { }
  {}
\theoremstyle{mystyle2}

\usepackage{hyperref}
\usepackage{xcolor}
\definecolor{rsrs}{RGB}{19, 59, 123}
\definecolor{myred}{rgb}{0.7, 0.0, 0.0}
\definecolor{mygreen}{rgb}{0.0, 0.3, 0.0}
\hypersetup{
    colorlinks,
    linkcolor={rsrs},
    citecolor={rsrs},
    urlcolor={rsrs}
}

\usepackage{tocloft}

\usepackage[nottoc]{tocbibind}

\usepackage{kantlipsum}

\usepackage[
    autocite=superscript,
    backend=biber,
    citestyle=chem-acs,
    bibstyle=numeric,
    natbib=true,
    maxbibnames=99,
    maxcitenames=2,
    giveninits=true,
    terseinits=true,
    uniquelist=false
]{biblatex}

\DefineBibliographyStrings{english}{%
    andothers = {\em et\addabbrvspace al.}
}
\DeclareNameAlias{author}{family-given}
\DeclareFieldFormat*{title}{#1}
\renewbibmacro{in:}{}

\citetrackerfalse
\DefineBibliographyStrings{english}{%
  mathesis = {MSc thesis},
}

\usepackage[title]{appendix}

\usepackage{authblk}

\usepackage[hang]{footmisc}
\usepackage[ruled, algo2e]{algorithm2e} 
\usepackage{algpseudocode}
\SetKw{KwInitialization}{Initialize}
\SetKw{KwInput}{Input:}
\SetKw{KwOutput}{Output:}

\usepackage{changepage}

\usepackage{booktabs}
\iftoggle{medphysjournal}{
    \usepackage{lineno}
    \linenumbers
}

\newcommand{\xs}[1]{x_{s_{#1}}}
\newcommand{\ys}[1]{y_{s_{#1}}}
\newcommand{\yhats}[1]{\hat{y}_{s_{#1}}}

\usepackage[noabbrev]{cleveref}

\title{Exjobb}
\begin{document}

\title{ {\sffamily Robust automated radiation therapy treatment planning using scenario-specific dose prediction and robust dose mimicking} }
\author[$\dagger$]{Oskar Eriksson\thanks{These authors contributed equally.}}
\newcommand\CoAuthorMark{\footnotemark[\arabic{footnote}]}
\author[$\dagger$\textsf{,}$\ddagger$]{Tianfang Zhang\protect\CoAuthorMark}
{
\affil[$\dagger$]{RaySearch Laboratories, Eugeniavägen 18, Solna, Stockholm SE-171 64, Sweden}
\affil[$\ddagger$]{Department of Mathematics, KTH Royal Institute of Technology, Stockholm SE-100 44, Sweden}
}
\iftoggle{medphysjournal}{
    \date{\vspace{-0.5cm }{\small Running title: Robust automated radiation therapy treatment planning} \\
    {\small Corresponding author: Tianfang Zhang}
    {\small E-mail: \url{tianfang@kth.se}} 
    {\small Address: RaySearch Laboratories, Eugeniavägen 18, Solna, Stockholm SE-171 64, Sweden} \\
    \vspace{0.2cm} \textsf{\today}}
}

\maketitle

\vspace{-1cm}

\begin{quote}
{\centering
\section*{Abstract}
}
\textit{Purpose}: We present a framework for robust automated treatment planning using machine learning, comprising scenario-specific dose prediction and robust dose mimicking. 

\noindent \textit{Methods}: The scenario dose prediction pipeline is divided into the prediction of nominal dose from input image and the prediction of scenario dose from nominal dose, each using a deep learning model with U-net architecture. By using a specially developed dose--volume histogram--based loss function, the predicted scenario doses are ensured sufficient target coverage despite the possibility of the training data being non-robust. Deliverable plans may then be created by solving a robust dose mimicking problem with the predictions as scenario-specific reference doses.

\noindent \textit{Results}: Numerical experiments are performed using a dataset of $52$ intensity-modulated proton therapy plans for prostate patients. We show that the predicted scenario doses resemble their respective ground truth well, in particular while having target coverage comparable to that of the nominal scenario. The deliverable plans produced by the subsequent robust dose mimicking were showed to be robust against the same scenario set considered for prediction.

\noindent \textit{Conclusions}: We demonstrate the feasibility and merits of the proposed methodology for incorporating robustness into automated treatment planning algorithms. 
\newline
\begin{spacing}{0.9}
{\sffamily\small \noindent \textbf{Keywords:} Knowledge-based planning, scenario dose prediction, robust optimization, dose mimicking.}
\end{spacing}
\end{quote}

\tolerance=1000

\newpage

\section{Introduction}

Radiation therapy treatment planning is a time-consuming process that typically requires multiple iterations between a dosimetrist and an oncologist \citep{feasibility}. In recent years, automated treatment planning methods have been developed to speed up the process while ensuring a consistent quality of treatment plans, using historically delivered treatment plans to aid in the process of creating plans for new patients. A common approach is using a machine learning model to predict a reference dose distribution for each new patient \citep{ge, siddique, ng}, which is then used in an optimization problem to find a configuration for a treatment machine that would deliver a similar dose \citep{atpprediction}. These parts are commonly referred to as dose prediction and dose mimicking, respectively. However, in many cases, especially in proton therapy, it is important to take into account uncertainties in the treatment delivery such as patient setup and density calculations \citep{minimaxopt}. While robust optimization is the current state-of-the-art method of handling such uncertainties, optimizing on e.g. the near-worst-case over a set of sampled scenarios, current methods for automated planning in the literature have yet to be able to incorporate robustness. The contribution of this paper is a framework that unifies ideas from automated treatment planning and robust optimization by predicting a set of scenario reference doses, each corresponding to a specific scenario, to be used in a robust dose mimicking problem.

Typically, for the non-robust case, machine learning--based automated treatment planning is divided into two steps: predicting the achievable values of certain dose-related quantities from the patient geometry, and finding machine parameters of a plan to reconstruct the same values \citep{ge}. For example, the dose-related quantities may be the spatial dose distribution, dose--volume histograms (DVHs), or a combination thereof. A common approach to spatial dose prediction is using a neural network, often with a U-net architecture, to predict a dose value for each voxel in the discretized dose grid \citep{dosenet, 3dunetdoseprediction, 3duresnet, uresnetd, hdunet, beamconfiguration}. Likewise, for DVH prediction, one may either use the DVHs evaluated on a spatially predicted dose distribution or employ separate models for the purpose \citep{appenzoller, jiao, wu, ma_features, yuan, zhang_dvhpred}. Using the reference dose and the reference DVHs, a dose mimicking optimization problem is then constructed, where the goal is to find a set of machine parameters for a specific treatment machine such that the delivered dose is as similar as possible to the reference dose and DVHs \citep{refdvh, dosemimicking, atpprediction}. Thus, given that the plans used for training the models are clinically acceptable and follow the desired protocol, the idea is that the automated planning pipeline should produce a high-quality plan for each new patient.

In robust optimization, reference dose levels are usually set for different regions of interest (ROIs), and the goal of the optimization is to find a resulting treatment plan that minimizes some cost functional that depends on the outcome in a number of fixed scenarios \citep{scenarios}. These scenarios may be seen as samples from some probability distribution specifying the geometric uncertainties in the treatment delivery that we want to account for. Typical examples of such uncertainties include setup uncertainties, related to inaccuracies in the setup of the patient or the inaccuracy of the treatment machine, as well as range uncertainties, related to uncertainties in the CT imaging or the conversion from CT values to density values \citep{minimaxopt}. For particle modalities such as protons, since the dose delivered is relatively sensitive to the density of the material through which they pass, using margins around the clinical target volume (CTV) is often insufficient to account for density uncertainty effects. Hence, robust optimization is especially important in such cases. In robust dose mimicking, one would use a reference dose corresponding to each scenario, with the goal of finding a robust treatment plan that is good across most or all of these scenarios \citep{robustmimicking}. However, in a robust prediction--mimicking pipeline, how to choose these scenario reference doses is a matter previously unaddressed in the literature.

In this paper, we present a method of performing robust automated treatment planning, combining spatial dose prediction and robust optimization through scenario dose prediction and robust dose mimicking. We propose to predict scenario doses using a two-step pipeline: by first predicting the nominal dose using a U-net model and subsequently deforming, using a second U-net model, the nominal dose to scenario doses corresponding to a given set of scenarios. Specifically, starting from a non-robustly planned dataset, we propose to use a DVH-based loss function when training the scenario model to ensure that the predicted scenario doses have comparable target coverage to that of the nominal dose. The predicted scenario doses are then used as scenario-specific reference doses in a robust dose mimicking problem, creating a robust deliverable plan. Numerical experiments, designed for a proof-of-concept study, are performed on a dataset of prostate cancer patients treated with intensity-modulated proton therapy. We show that the proposed scenario dose prediction pipeline fulfills its purpose satisfactorily, with predictions mostly following the non-robust ground truth but with increased target coverage, and that the resulting deliverable plans are robust with respect to the considered scenario set. Compared to manually generated benchmark plans, the automatically generated plans are similar in terms of DVH, dose statistic spread, and spatial dose. In particular, we demonstrate the feasibility of a data-driven, robust automated treatment planning framework.

\section{Materials and methods}

Let $\mathcal{X}$ and $\mathcal{Y}$ be spaces of patient geometries and dose distributions, respectively, and $\mathcal{S}$ a given set of scenarios representing realizations of systematic setup or range uncertainties. For each $s \in \mathcal{S}$, the patient geometry $x_s \in \mathcal{X}$ is defined as its CT image with ROI delineations along with its spatial location under $s$. A dose distribution that has been delivered to this patient previously is referred to as the scenario dose, denoted by $y_s$. The nominal scenario, corresponding to no setup or range errors, is denoted by $s_0$. Given a dataset $\{(\xs{0}^{n}, \ys{0}^{n})\}_{n=1}^N \subset \mathcal{X} \times \mathcal{Y}$ of pairs of nominal patient geometries and doses, our proposed pipeline follows the classical prediction--mimicking division, with a scenario-specific spatial dose prediction model followed by a robust dose mimicking optimization.

\subsection{Pipeline}

\begin{figure}[htbp]
\centering
\includegraphics[scale=0.7]{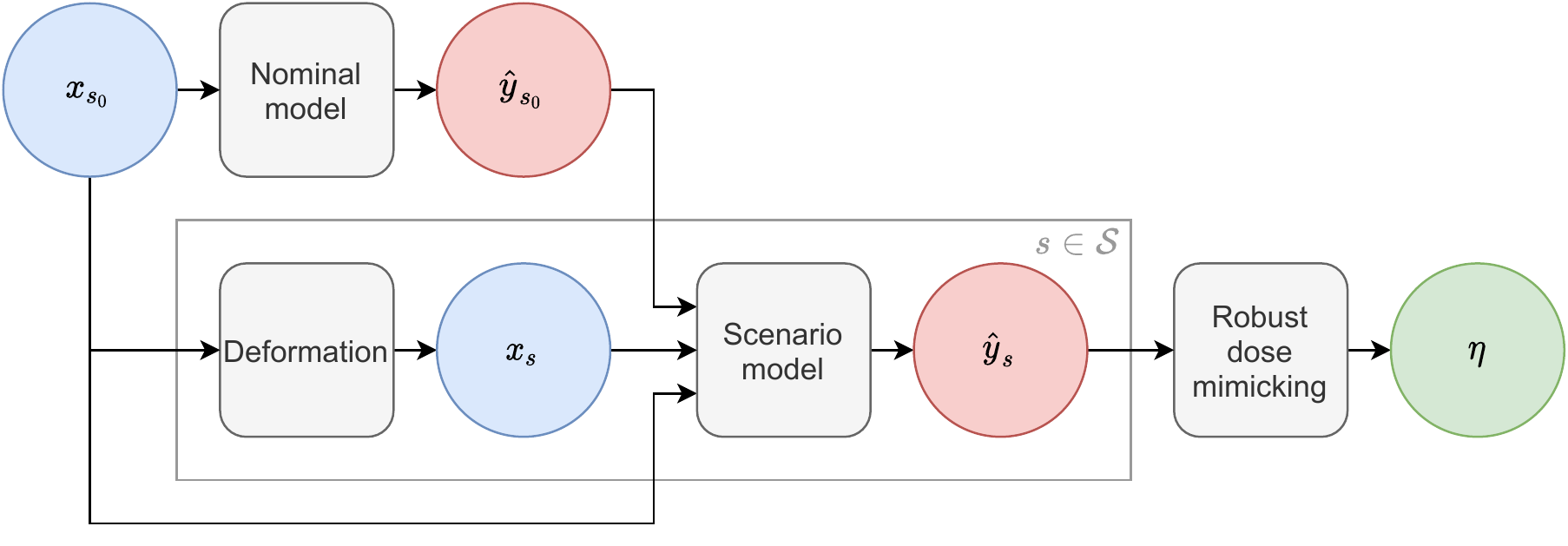}
\caption{An illustration of the proposed pipeline. A nominal dose $\yhats{0}$ is predicted from a nominal geometry $\xs{0}$ using a nominal model. Furthermore, for each scenario $s \in \mathcal{S}$, $\xs{0}$ is deformed into a scenario geometry $x_s$. The scenario model is then used to predict a scenario dose $\hat{y}_s$ from these for each scenario. The scenario doses are finally used in robust dose mimicking yielding a set of machine parameters $\eta$.}
\label{fig:robust-pipeline}
\end{figure}

We propose to extend the typical nominal spatial dose prediction and dose mimicking procedure with a scenario-specific dose prediction component, as illustrated in \Cref{fig:robust-pipeline}. First, a nominal model predicts the nominal dose $\yhats{0} \in \mathcal{Y}$ from the nominal geometry $\xs{0}$. To predict the scenario dose $\hat{y}_s$ for any scenario $s \in \mathcal{S}$, $\yhats{0}$ is deformed in accordance with the change from the nominal geometry $\xs{0}$ to the scenario geometry $x_s$ using a second scenario model. One such scenario dose is predicted for each $s\in\mathcal{S}$, yielding a set $\left\{\hat{y}_s\right\}_{s\in\mathcal{S}}$ of predicted scenario doses.

\begin{figure}[htbp]
\centering
\includegraphics[scale=0.7]{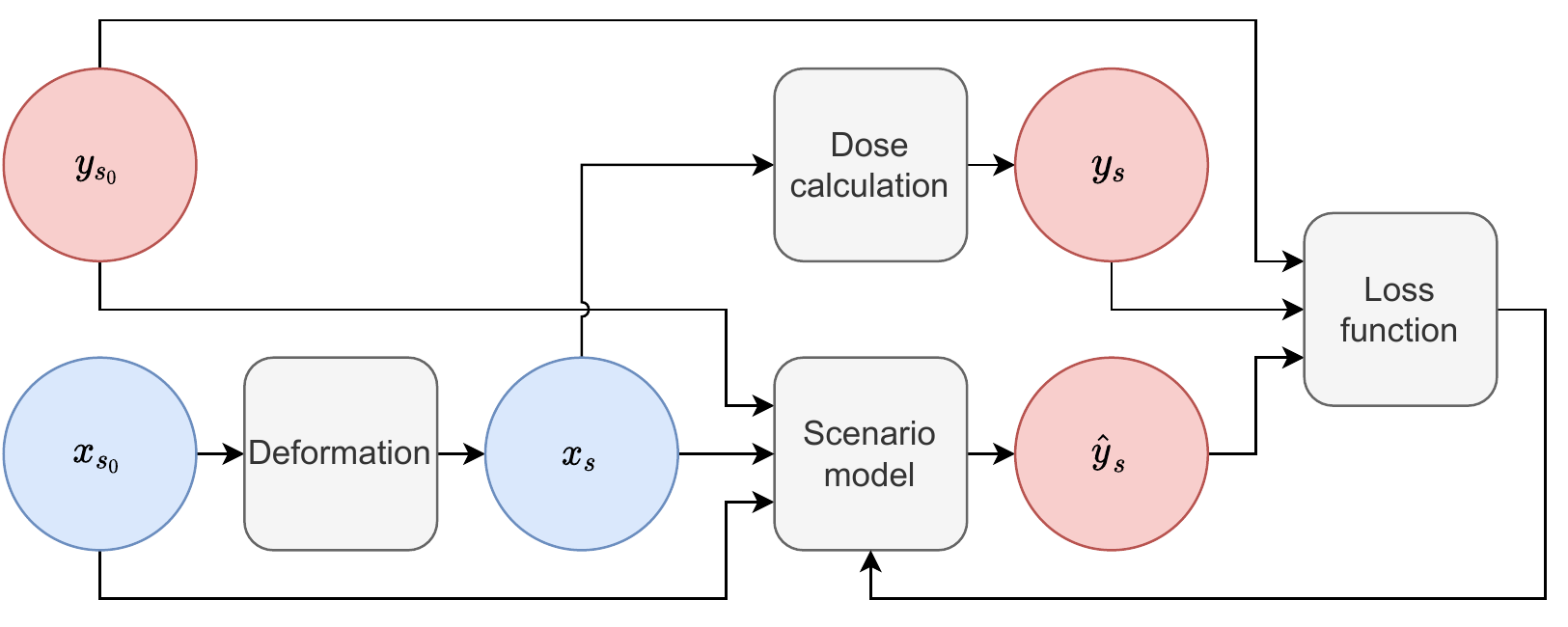}
\caption{An illustration of training the proposed pipeline. For each scenario $s$, the ground truth dose $y_s$ is calculated using an accurate dose calculation algorithm from the corresponding scenario geometry $x_s$. The scenario model is trained using a loss function consisting of a spatial loss that depends on the predicted scenario dose $\hat{y}_s$ and $y_s$, as well as a DVH loss that depends on $\hat{y}_s$ and $\ys{0}$.}
\label{fig:training-pipeline}
\end{figure}

To train such a scenario model, ideally, the training data would consist of robust treatment plans---in particular, a set $\{(x_s^{n}, y_s^{n})\}_{n, \, s \in \mathcal{S}}$ of geometry--dose pairs where all doses are planned robustly with respect to the scenario set $\mathcal{S}$. However, to remove the need of having access to a complete dataset of robust plans, which is a relatively strict requirement, we propose an alternative method of training the scenario model. In this framework, each plan in a dataset of previously delivered non-robust treatment plans is deformed in accordance with a number of scenarios. As we want the resulting robust treatment plans to have sufficient target coverage, we want our scenario reference doses to have sufficient target coverage as well. Therefore, we propose to train the scenario model using a loss function that both penalizes deviation from the non-robust ground truth scenario dose, as well as deviation from the target coverage of the nominal dose. The training pipeline is illustrated in \Cref{fig:training-pipeline}.

\subsection{Algorithm}
\label{sec:algorithm}

To maintain the target coverage equivalent to that of the nominal dose, but still predicting a realistic scenario dose, we propose to train the scenario model using a loss function that combines a voxel-level spatial loss $L_{\operatorname{spat}}(\hat{y}, y) : \mathcal{Y}^2 \to \mathbb{R}$ with a DVH loss $L_{\operatorname{DVH}}(\hat{y}, y) : \mathcal{Y}^2 \to \mathbb{R}$ defined on the targets. In particular, the spatial loss is given by the weighted mean-squared error 
\begin{equation}
\label{eq:spatial-loss}
L_{\operatorname{spat}}(\hat{y}, y) = \sum_i \omega_i (\hat{y}_i - y_i)^2,
\end{equation}
where the $\omega_i$ are nonnegative voxel weights such that $\sum_i \omega_i = 1$. For the DVH loss, let $\mathcal{R}$ be the set of all ROIs, each $R \in \mathcal{R}$ represented as index sets of voxels, and let $y_R = (y_i)_{i \in R}$ be the local dose vector for each $R$. Let, furthermore $\mathcal{R}_{\operatorname{target}} \subseteq \mathcal{R}$ be the set of all target ROIs which are to be covered robustly, typically chosen to comprise all CTVs. Denoting by $\operatorname{D}_v$ the dose-at-volume at volume level $0 \leq v \leq 1$, which is given for each $R$ by 
\[
\operatorname{D}_v(y_R) = \inf\!\left\{ x \in \mathbb{R} : \frac{1}{|R|} \sum_{i \in R} 1_{y_i \; \geq \; x} \leq v \right\}
,\]
the DVH loss may be written as
\[
L_{\operatorname{DVH}}(\hat{y}, y) = \sum_{R \in \mathcal{R}_{\operatorname{target}}} \int_0^1 \left( \operatorname{D}_v(\hat{y}_R) - \operatorname{D}_v(y_R) \right)^2 dv
.\]
The idea of utilizing a DVH loss for training dose prediction models has previously been explored by \citet{nguyen_pareto} and \citet{zhang_mco}, but for other purposes. Upon predicting $\hat{y}_s$ in a scenario $s \in \mathcal{S}$ with non-robust scenario ground truth $y_s$ and nominal ground truth $\ys{0}$, the loss contribution is then given by 
\[
L(\hat{y}_s, y_s, \ys{0}) = L_{\operatorname{spat}}(\hat{y}_s, y_s) + \alpha L_{\operatorname{DVH}}(\hat{y}_s, \ys{0})
,\]
weighting the DVH loss by a factor $\alpha$. 

\begin{figure}[htbp]
\centering
\includegraphics[width=\textwidth]{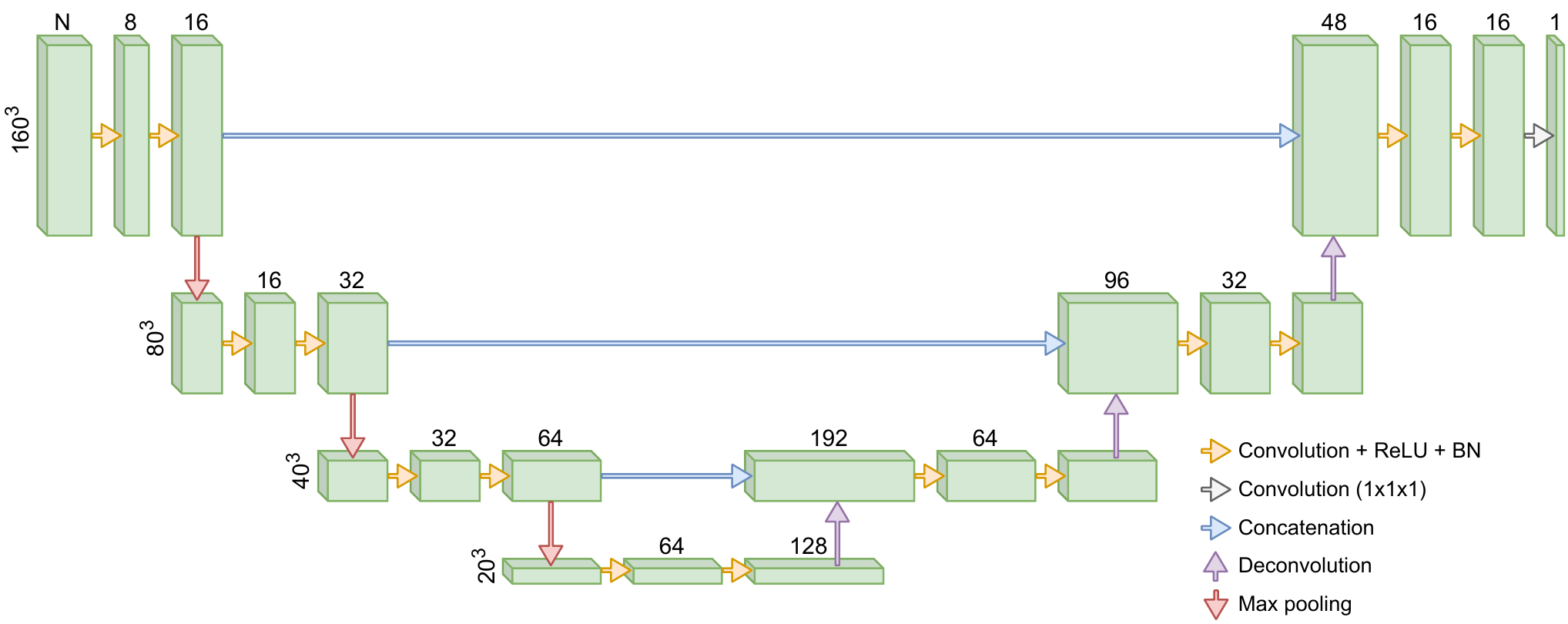}
\caption{The proposed 3D U-net architecture. The size of each row is denoted at the far left and the number of channels in each layer is denoted above the layer.}
\label{fig:architecture}
\end{figure}

For both the nominal and scenario models, we propose to use an architecture based on the 3D U-net \citep{3dunet}, with dimensions adapted for the present dose prediction purpose. The architecture used in our experiments is illustrated in \Cref{fig:architecture}. This architecture consists of a contracting path, consisting of a series of convolution, activation, batch normalization, and max pooling operations, serving as a form of feature extraction, and a symmetric expanding path that instead uses deconvolution operations, with skip connections between the two paths. Similar architectures have previously been used successfully for the task of dose prediction \citep{dosenet, 3dunetdoseprediction, 3duresnet, uresnetd, hdunet, beamconfiguration}.

\subsection{Robust dose mimicking}

Having trained the nominal and scenario models to obtain for a new patient a set $\{\hat{y}_s\}_{s \in \mathcal{S}}$ of predictions, we may use the same predictions as scenario-specific reference doses in a dose mimicking to create a robust deliverable plan. Let $\eta \in \mathcal{E}$ denote the machine parameters, the physical limitations of which are articulated by constraining $\eta$ to lie in the feasible set $\mathcal{E}$, and let $\{y_s(\eta)\}_{s \in \mathcal{S}}$ be the scenario doses resulting from the plan defined by $\eta$. Partitioning $\mathcal{R} = \mathcal{R}_{\operatorname{non-robust}} \cup \mathcal{R}_{\operatorname{robust}}$ into non-robust and robust ROIs, we can write a general robust dose mimicking problem as 
\[
\underrel{\text{minimize}}{\eta \in \mathcal{E}} \quad \sum_{R \in \mathcal{R}_{\operatorname{non-robust}}} \psi_{\operatorname{mimic}, R}(y_{s_0}(\eta), \hat{y}_{s_0}) +  C\!\left( \left( \sum_{R \in \mathcal{R}_{\operatorname{robust}}} \psi_{\operatorname{mimic}, R}(y_s(\eta), \hat{y}_s) \right)_{s \in \mathcal{S}} \right)\!,
\]
where $C : \mathbb{R}^{|\mathcal{S}|} \to \mathbb{R}$ is a cost functional expressing the conservativeness of the robust optimization and each $\psi_{\operatorname{mimic}, R} : \mathcal{Y}^2 \to \mathbb{R}$ is on the form 
\[
\psi_{\operatorname{mimic}, R}(y, \hat{y}) = w_{\operatorname{spat}, R} \psi_{\operatorname{spat}}(y_R, \hat{y}_R) + w_{\operatorname{DVH}, R} \psi_{\operatorname{DVH}}(y_R, \hat{y}_R).    
\]
Here, the cost functional is commonly chosen as a weighted power mean function $C( (\xi_s)_{s \in \mathcal{S}}) = \left( \sum_{s \in S} \xi^a \right)^{1 / a}$ with exponent $a \geq 1$, where $a = 1$ corresponds to stochastic programming and $a = \infty$ to minimax optimization \citep{scenarios}. As for $\psi_{\operatorname{spat}}$ and $\psi_{\operatorname{DVH}}$, common choices are one-sided quadratic functions penalizing deviations in spatial dose and DVH, respectively.

\subsection{Computational study}

We applied our framework to a cohort of $52$ retrospective treatment plans for prostate cancer patients, divided into a training set of $37$ patients, a test set of $10$ patients, and a validation set of $5$ patients. The patients originally received volumetric modulated arc therapy, so we utilized the framework proposed by \citet{robustmimicking} to convert them to proton pencil beam scanning (PBS) plans. We used two beams with isocenters in the middle of the CTV, i.e. the prostate, aimed at $90$ and $270$ degrees, going through the left and right femurs respectively. The organs at risk (OARs) considered were the rectum, bladder, anal canal, left and right femurs. The dose prescribed to all the patients was a median dose of $7 700 \; \mathrm{cGy}$ ($\operatorname{RBE} = 1.1$) to the CTV, delivered over $35$ fractions.

For the uncertainties, similar to \citet{minimaxopt}, we used a density uncertainty of $\pm 3 \; \%$, and a slightly smaller setup uncertainty of $0.5 \; \mathrm{cm}$. For the setup uncertainty, the isocenter was shifted in the unit directions as well as diagonally. The identity shift was also included for both the density and setup uncertainties, resulting in a total of $|\mathcal{S}| = 45$ scenarios with corresponding geometries $\{x_s\}_{s \in \mathcal{S}}$ for each patient. For both the nominal and scenario models, we used the architecture in \Cref{fig:architecture}, with sizes and channels adjusted for our purposes. The voxel size used was $0.25 \times 0.25 \times 0.25 \; \mathrm{cm}^3$, and the input and output resolution of the models were $160 \times 160 \times 160$ voxels. For the nominal model, the input channels consisted of $\xs{0}$, i.e. the binary masks for the CTV and each OAR as well as a CT image of the patient. Recall that the purpose of the scenario model is to deform the nominal dose in accordance with a specific scenario $s$---therefore, the scenario model had input channels consisting of both the nominal geometry $\xs{0}$ and a scenario geometry $x_{s}$, as well as the nominal dose $\ys{0}$.

Both models were trained using the total loss described in \Cref{sec:algorithm}. For the scenario model, we used the training dataset of size $N |\mathcal{S}| = 37 \cdot 45 = 1665$ resulting from including all scenarios for each patient. In both the nominal and scenario models, data augmentation was applied in the form of rotations around the transversal axis, drawn uniformly at random between $-10^{\circ}$ and $10^{\circ}$ at each forward pass through the network. The DVH loss weighting factor $\alpha$ was chosen using grid search over the values in $\left\{ 0, 0.1, 0.2, 0.3, 0.4, 0.5 \right\}$. The models were trained until the loss on the validation set converged, which was around $500$ epochs for the nominal model and $50$ epochs for the scenario model. To select the weighting factor $\alpha$, we inspected the dose washes and DVH curves. For both models, depending on the value of $\alpha$, we saw a tradeoff between the ability to predict a dose that gives a good DVH for the CTV, and a realistic dose outside of the CTV. For the nominal and the scenario model, we used $\alpha=0.1$ and $\alpha = 0.3$, respectively. We observed that these weighting factors gave a satisfactory CTV coverage, while still giving a realistic dose outside of the CTV. The spatial loss $L_{\operatorname{spat}}$ in (\ref{eq:spatial-loss}) chosen for our experiments is a voxel-level mean squared error loss. As voxels closer to the CTV are generally more important, we weighted the contribution of each voxel depending on its distance from the CTV using the weighting $\omega_i \propto \max\{ e^{-\beta \operatorname{DT}_{\operatorname{CTV}}(i)}, \omega_{\min}\}$, where $\operatorname{DT}_{\operatorname{CTV}}(i)$ is the Euclidean distance from the voxel $i$ to the CTV, $\beta > 0$ is a constant and $\omega_{\min}$ is a minimum weight threshold. For our experiments, we used $\beta = 0.05$ and $\omega_{\min} = 0.01$.

The robust dose mimicking was performed using a research version of the treatment planning system RayStation 11A (RaySearch Laboratories, Stockholm, Sweden) with sequential quadratic programming--based optimization, creating deliverable proton PBS plans. The in-house dose mimicking algorithm was used, with one-sided quadratic loss functions $\psi_{\operatorname{spat}}$ and $\psi_{\operatorname{DVH}}$, voxel-level weights determined partly depending on the corresponding isodose level on the nominal predicted dose, and a weighted-power-mean cost functional with exponent $8$ approximating minimax optimization. The ROIs $\mathcal{R}$ considered in the dose mimicking problem were the CTV, bladder, rectum, and ring ROIs with distances $0\mathrm{-}1 \; \mathrm{cm}$ cm and $1\mathrm{-}2 \; \mathrm{cm}$ cm from the CTV, all included in the robust subset $\mathcal{R}_{\mathrm{robust}}$. The mimicking optimization was divided into three runs of $60$, $60$, and $8$ iterations respectively. In particular, approximate doses during optimization were computed using a Monte Carlo algorithm using $10^4$ ions per spot, and final doses using the same algorithm with a statistical uncertainty of $0.5 \; \%$. 

Finally, the plans produced by the proposed automated algorithms were benchmarked to manually generated robust treatment plans for the same patients. The in-house inverse planning algorithm in RayStation 11A was employed to create the comparison plans using the optimization functions displayed in \Cref{tab:manual_objectives}, of which all were robust and those on the CTV were formulated as constraints. In particular, the optimization was run using the same weighted-power-mean exponent, number of iterations and dose computation settings as for the automatically generated plans. While additional fine-tuning of the manual plans are likely to be needed before reaching fully clinical quality, they serve the purpose of a comprehensive and transparent benchmark for the proposed automated planning algorithm.

\begin{table}[h]
\caption{\label{tab:manual_objectives}The optimization formulation used for creating the manual comparison plans.}
\centering
\begin{tabular}{lllll}
\toprule
ROI & Optimization function & Robust & Constraint & Weight \\
\midrule
CTV & At least $7400 \; \mathrm{cGy}$ at $98 \; \%$ volume & Yes & Yes & -- \\
CTV & At most $7900 \; \mathrm{cGy}$ at $2 \; \%$ volume & Yes & Yes & -- \\
Bladder & At most $6500 \; \mathrm{cGy}$ at $10 \; \%$ volume & Yes & No & $1$ \\
Rectum & At most $6500 \; \mathrm{cGy}$ at $10 \; \%$ volume & Yes & No & $1$ \\
Ring $0\mathrm{-}1 \; \mathrm{cm}$ & At most $7000 \; \mathrm{cGy}$ mean dose & Yes & No & $1$ \\
Ring $0\mathrm{-}1 \; \mathrm{cm}$ & At most $4000 \; \mathrm{cGy}$ mean dose & Yes & No & $1$ \\
\bottomrule
\end{tabular}
\label{manual_objectives}
\end{table}

\section{Results}

\begin{figure}[htbp]
\centering
\includegraphics[width=0.7\linewidth]{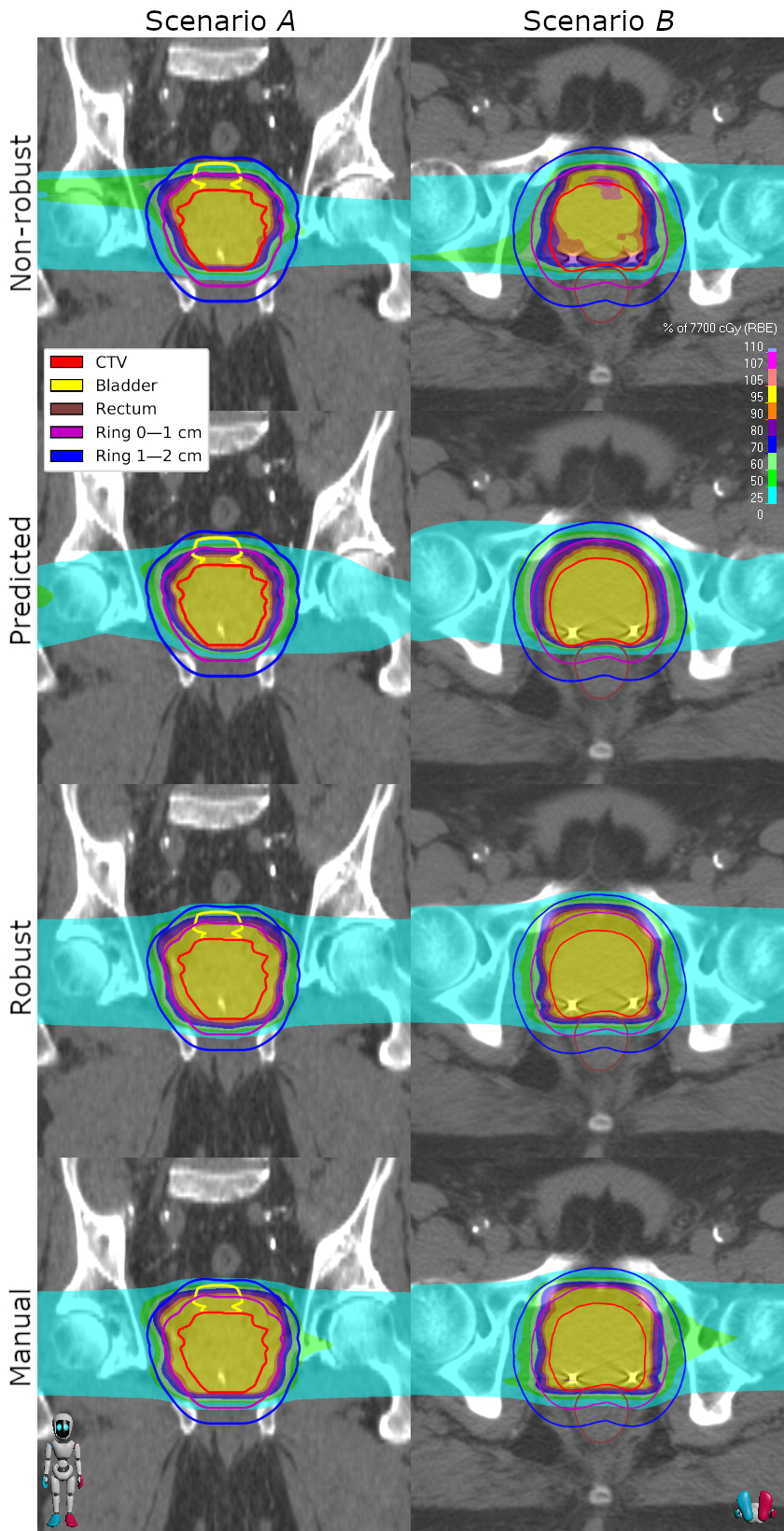}
\caption{Two different scenarios with the non-robust (first row), the predicted (second row), the robust (third row), and the manual (fourth row) doses. The left column shows a coronal slice of scenario $A$ where the patient has been translated down $0.5 \; \mathrm{cm}$ with respect to the image. The right column shows a transversal slice of scenario $B$ where the patient has been translated down $0.5 \; \mathrm{cm}$ with respect to the image and a density shift of $+ 3 \; \%$ has been applied.}
\label{fig:spatial-plots}
\end{figure}

\begin{figure}[htbp]
\centering
\includegraphics[width=\linewidth]{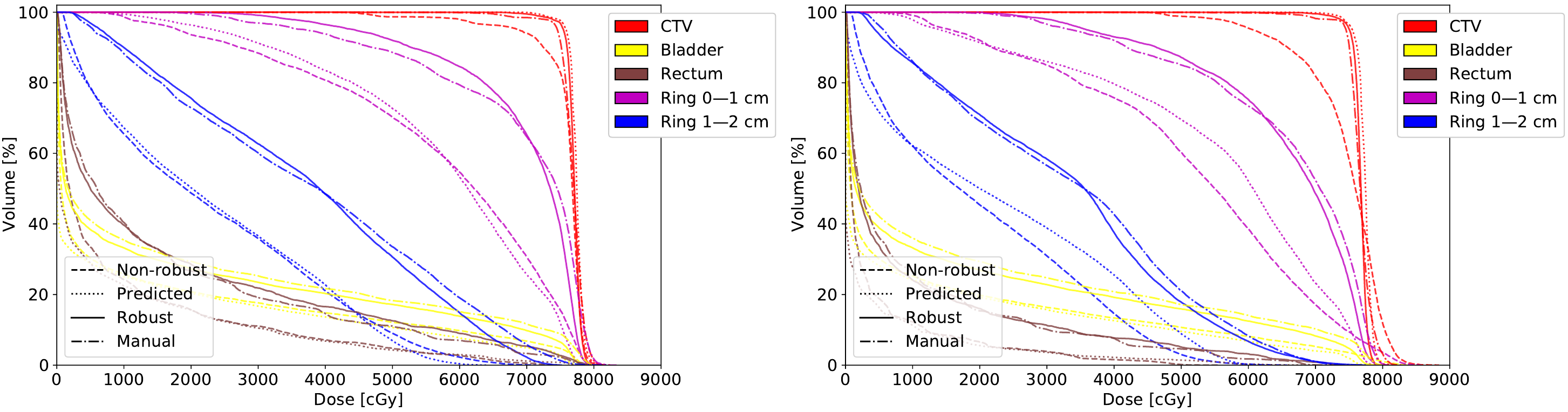}
\caption{The DVHs for scenario $A$ (left) and scenario $B$ (right).}
\label{fig:dvhs}
\end{figure}

To verify the feasibility of the proposed pipeline, we perform a qualitative analysis for one test patient and a quantitative analysis based on the entire test dataset. In \Cref{fig:spatial-plots}, two scenarios for a test patient are visualized. The non-robust ground truth dose fails to give a sufficient target coverage---in particular, in the transversal view, one can see that large parts of the CTV receive less than $95 \; \%$ of the prescribed dose. However, the scenario dose prediction model has been trained to predict doses with a better target coverage, and we see indeed that the predicted dose successfully covers the CTV in these scenarios. Finally, the automatically generated robust deliverable dose is generated by performing a robust dose mimicking using all the predicted scenario doses as reference doses and is expected to ensure coverage of the CTV in most or all of the scenarios. For the two scenarios displayed, the CTVs are well-covered, whereas the dose decays slower beyond the CTV outline than the corresponding predicted scenario doses---this is due to the robust plan needing to account for the outcome in all scenarios. The manually generated benchmark has similar target coverage and decay around the CTV compared to the automatically generated dose, but slightly more dose spillage in the area beyond 2 cm from the CTV.

Furthermore, the DVHs corresponding to the scenarios in \Cref{fig:spatial-plots} are displayed in \Cref{fig:dvhs}. The DVHs of the predicted scenario doses are relatively similar to those of the non-robust dose in all ROIs except for the CTV, where the target coverage is instead more similar to that of the prescribed dose. This is what we wanted to achieve with our scenario dose prediction since the spatial component of the loss function used to train the scenario model is expected to make the predicted doses similar to the ground truth scenario doses, while the DVH component is aimed at maintaining the target coverage of the nominal dose. We can also see that the automatically generated robust dose has a similar target coverage as the predicted dose in both scenarios, meaning that the target coverage of the predicted doses successfully propagates to the robust dose---however, naturally, this comes with the cost of a slight increase in dosage to the rectum, bladder and the target surroundings. The automatically generated robust dose has similar DVHs compared to those of the manual benchmark. 

\begin{figure}[htbp]
\centering
\includegraphics[width=\linewidth]{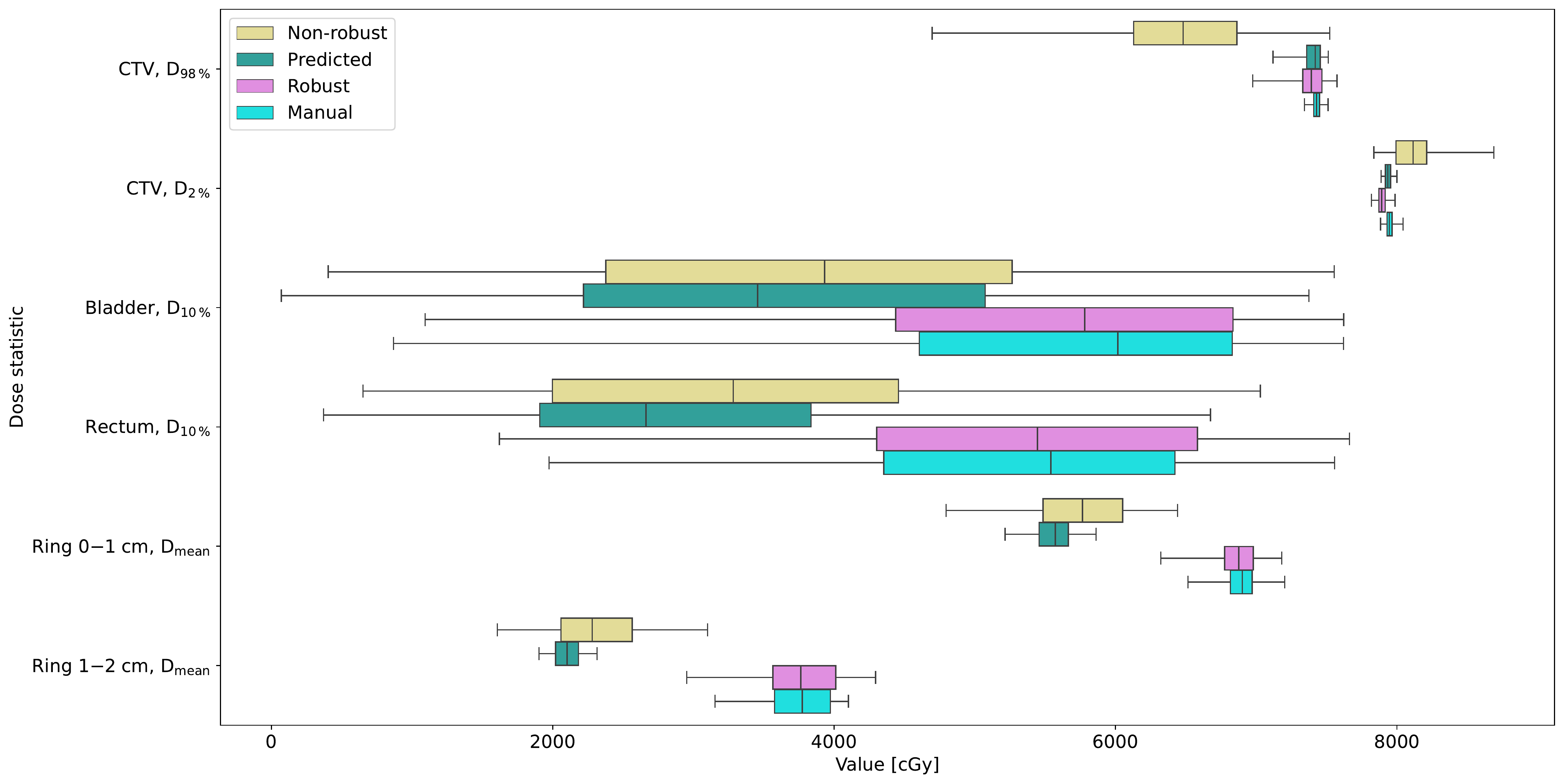}
\caption{Boxplot of dose statistics for the different dose types evaluated across the 45 scenarios for each test patient.}
\label{fig:boxplot}
\end{figure}

\begin{table}[htbp]
\caption{The minimum, maximum, mean, and standard deviation of the dose statistics for the different dose types evaluated across the $45$ scenarios for each test patient.}
\label{tab:ds}
\centering
\begin{tabular}{llrrrr}
\toprule
Goal & Type &         Min ($\mathrm{cGy}$) &         Max ($\mathrm{cGy}$) &        Mean ($\mathrm{cGy}$) &         Std ($\mathrm{cGy}$) \\
\midrule
& Non-robust &  4697 & 7523 & 6488 &  542 \\
CTV, $\operatorname{D}_{98\, \%}$ &  Predicted &  7119 & 7512 & 7401 &   73 \\
&     Robust &  6975 & 7574 & 7386 &  111 \\
&   Manual &  7342 & 7511 & 7429 &   31 \\
\midrule
& Non-robust &  7836 & 8688 & 8117 &  159 \\
CTV, $\operatorname{D}_{2\,\%}$ &  Predicted &  7887 & 8000 & 7939 &   25 \\
&     Robust &  7819 & 7987 & 7894 &   31 \\
&   Manual &  7884 & 8044 & 7948 &   28 \\
\midrule
& Non-robust &   404 & 7555 & 3885 & 1793 \\
Bladder, $\operatorname{D}_{10\,\%}$ &  Predicted &    70 & 7375 & 3500 & 1869 \\
&     Robust &  1093 & 7622 & 5481 & 1648 \\
&   Manual &   868 & 7620 & 5495 & 1726 \\
\midrule
& Non-robust &   651 & 7029 & 3338 & 1586 \\
Rectum, $\operatorname{D}_{10\,\%}$ &  Predicted &   371 & 6675 & 2947 & 1524 \\
&     Robust &  1620 & 7662 & 5362 & 1462 \\
&   Manual &  1974 & 7557 & 5349 & 1316 \\
\midrule
& Non-robust &  4795 & 6440 & 5742 &  369 \\
Ring $0\mathrm{-}1 \; \mathrm{cm}$, $\operatorname{D}_\mathrm{mean}$ &  Predicted &  5216 & 5862 & 5565 &  136 \\
&     Robust &  6321 & 7181 & 6867 &  157 \\
&   Manual &  6515 & 7202 & 6892 &  136 \\
\midrule
& Non-robust &  1606 & 3101 & 2297 &  350 \\
Ring $1\mathrm{-}2 \; \mathrm{cm}$, $\operatorname{D}_\mathrm{mean}$ &  Predicted &  1902 & 2315 & 2104 &  104 \\
&     Robust &  2952 & 4294 & 3755 &  306 \\
&   Manual &  3154 & 4102 & 3746 &  244 \\
\bottomrule
\end{tabular}
\end{table}

Moreover, in \Cref{fig:boxplot} and \Cref{tab:ds}, a number of dose statistics aggregated across all test patients and scenarios are presented. For CTV $\operatorname{D}_{98\,\%}$ and $\operatorname{D}_{2\,\%}$, we see that the predicted and automatically generated robust doses have a lower spread and are more focused around the prescribed dose than the non-robust dose, whereas they have a slightly higher spread compared to the manual plans. This indicates that the automatically generated dose is in fact robust with respect to the CTV given the specified uncertainty parameters. For bladder and rectum $\operatorname{D}_{10\,\%}$, we see that the automatically generated dose gives a higher dosage in general to the OARs than the non-robust and predicted doses, which is an expected effect of delivering more dose to the area around the CTV due to the overlap between these OARs and the CTV in the different scenarios---however, the dosage is similar to that of the manual plans. Finally, we include two ring ROIs, $0\mathrm{-}1 \; \mathrm{cm}$ representing a border of $1 \; \mathrm{cm}$ around the CTV and $1\mathrm{-}2 \; \mathrm{cm}$ representing a border of $1 \; \mathrm{cm}$ around the aforementioned ring ROI, with the purpose of illustrating the decay of dose beyond the target. We can see here, as well as in Figures \ref{fig:spatial-plots} and \ref{fig:dvhs}, that the decrease is slower for the automatically generated dose than for the non-robust and predicted doses, which is an expected effect of delivering more of the prescribed dose. However, the decrease is very similar to that of the manually generated dose. In summary, the proposed pipeline is able to generate doses that are robust with respect to the scenarios, while we see certain expected effects from delivering more dose than in the non-robust case.

\section{Discussion}

In this work, we have presented a data-driven approach to robust automated radiation therapy treatment planning. Using a dataset of non-robust proton plans and a two-step scenario dose prediction model with a DVH-based loss function term, we were able to predict relatively realistic scenario doses with target coverage comparable to that of the nominal ground truth dose. By using them as scenario-specific reference doses in a robust dose mimicking problem, we were able to create robust deliverable plans consistently delivering sufficient target coverage at the cost of an expectedly higher dosage to surrounding tissue than the predicted and non-robust doses. Compared to manually generated benchmark plans, the produced plans were similar in terms of DVH, dose statistic spread and spatial dose. While additional post-processing of the automatically generated plans may be needed in order to be considered clinical, as is the case with automated planning algorithms in general, the results serve to showcase the feasibility of our type of workflow---that is, the combination of scenario dose prediction and robust dose mimicking.

Among the advantages of our method are the non-requirement of a robustly planned dataset for training, the generality and flexibility associated with separating the task of scenario dose prediction into a nominal and scenario model, and the more rigorous handling of setup and range uncertainties through robust optimization rather than through, for example, margins. Although experiments were performed on proton plans, for photons, especially, one may want to have the choice of whether or not to use robust planning. In such cases, access to a completely robust training dataset may be too high a requirement. By separating the dose prediction pipeline into the image-to-nominal and nominal-to-scenario parts, we may use instead use data obtained from robust evaluation to learn the physical deformations associated with each scenario. Along with the DVH loss aimed at controlling target coverage of the scenario model outputs, the result is a highly general framework for handling robustness in automated treatment planning. 

However, one disadvantage of the method is that there is a lack of theoretical rigor for predicting realistic scenario doses with better target coverage. With our loss functions, the optimal output of the scenario model is a dose that is identical to the non-robust ground truth outside the CTV and identical to the nominal dose within the CTV, which is a discontinuous dose. In practice, smoothness is introduced by the convolutional design of the neural network, but this notion is hard to control exactly. Insofar as the predicted scenario doses are similar to the non-robust scenario ground truth far from the target, similar to the nominal ground truth in the target and some smoothened mixture in between, they can be understood to represent a theoretically ideal scenario dose given a fixed nominal dose. They should, however, be strictly more realistic than the nominal dose in each non-nominal scenario---indeed, using the nominal dose as reference dose in all scenarios is guaranteed not to be achievable. Hence, even though the predicted scenario doses may not always be physically realizable, the introduction of an additional scenario-specific dose prediction shrinks the gap between reference dose and physically realizable dose in the dose mimicking phase. 

Apart from addressing the smoothness issue, future research may include evaluating the proposed methodology on datasets with different treatment modalities, delivery techniques, and robustness parameters. For protons and heavy ions, one may also generalize the current scenario dose prediction to a version in which beam doses are predicted separately, with the addition of beam-specific objective functions in the dose mimicking optimization problem. The scenario dose prediction may also be used in other contexts than automated planning, e.g. for quality assurance purposes. Moreover, one may try to combine the current framework with a semiautomatic multicriteria optimization methodology such as in \citet{zhang_mco}. All in all, the incorporation of robustness into machine learning--automated treatment planning enables ample new opportunities to be explored.

\section{Conclusions}

We have presented a new data-driven approach to robust automated treatment planning, combining prediction of spatial scenario doses with robust dose mimicking. By dividing the former part into a nominal and a scenario dose prediction model, and using a DVH loss during training, we are able to predict for each new patient scenario doses with a robustly covered target using a non-robust training dataset. Subsequently, through robust dose mimicking, we obtain plans robust against the same scenario set. The numerical results serve to demonstrate the feasibility of the proposed methodology, which has the potential of facilitating the incorporation of robustness into automated planning.

\section*{Acknowledgments}

The authors thank Mats Holmström for help with setting up the dose mimicking optimization problem in RayStation, and for fruitful discussions along with Fredrik Löfman.

\iftoggle{medphysjournal}{
    \section*{Conflict of interest}
    
    The authors have no conflicts to disclose.
    
    \section*{Data availability statement}
    
    Research data are not shared.
}

\printbibliography

\end{document}